\documentclass[sigplan]{acmart}

\usepackage{subcaption}
\usepackage{graphicx}
\usepackage{amsfonts}
\usepackage{multirow}
\usepackage{url}
\usepackage{listings}
\usepackage{tikz}
\usetikzlibrary{matrix}

\usetikzlibrary{shapes.geometric}
\tikzstyle{every node}=[font=\footnotesize]
\tikzstyle{etichetta}=[]
\tikzstyle{square}=[draw,outer sep=0pt,inner sep=-1.3em,regular polygon,regular polygon sides=4,,minimum size=13.5mm]
\tikzstyle{square2}=[square,dotted,black!50]
\pgfmathsetmacro{\ra}{0.95}
\usetikzlibrary{decorations.pathreplacing,calc}

\usepackage{listings}
\lstdefinelanguage{json}{
    showspaces=false,
    showtabs=false,
    breaklines=false,
    breakatwhitespace=false,
    basicstyle=\ttfamily\footnotesize,
    upquote=true,
    morestring=[b]",
    literate={♭}{$\flat$}1
}
\usepackage{algorithm}
\usepackage{algpseudocode}

\algnewcommand\algorithmicswitch{\textbf{switch}}
\algnewcommand\algorithmiccase{\textbf{case}}
\algnewcommand\algorithmicdefault{\textbf{default}}
\algnewcommand\algorithmicassert{\texttt{assert}}
\algnewcommand\Assert[1]{\State \algorithmicassert(#1)}%
\algdef{SE}[SWITCH]{Switch}{EndSwitch}[1]{\algorithmicswitch\ #1\ \algorithmicdo}{\algorithmicend\ \algorithmicswitch}%
\algdef{SE}[CASE]{Case}{EndCase}[1]{\algorithmiccase\ #1}{\algorithmicend\ \algorithmiccase}%
\algdef{SE}[DEFAULT]{Default}{EndCase}[1]{\algorithmicdefault\ #1}{\algorithmicend\ \algorithmiccase}%
\algtext*{EndSwitch}%
\algtext*{EndCase}%
\algnewcommand\algorithmicforeach{\textbf{for each}}
\algdef{S}[FOR]{ForEach}[1]{\algorithmicforeach\ #1\ \algorithmicdo}

\renewcommand\footnotetextcopyrightpermission[1]{}
\acmConference[]{}{}

\begin{document}

\title{Introducing Support for \\Move Operations in Melda CRDT}

\author{Amos Brocco}
\email{amos.brocco@supsi.ch}
\orcid{0000-0002-0262-2044}
\affiliation{%
  \institution{University of Applied Sciences and Arts of Southern Switzerland}
  \city{Lugano}
  \country{Switzerland}
}

\begin{abstract}
In this paper, we present an extension to Melda (a library which implements a general purpose delta state JSON CRDT) to support move operations. This enhancement relies on minimal changes to the underlying logic of the data structure, has virtually no runtime overhead and zero storage overhead compared to the original version of the library, ensuring simplicity while addressing multiple use cases. Although concurrent reordering of the elements in a list was already supported in the original version of the library, moving objects between different containers lead to undesired outcomes, namely duplicate entries. To address this problem we revisited the original approach and introduced the necessary changes to support for relocating elements within a JSON structure. We detail those changes and provide some examples.
\end{abstract}

\maketitle

\section{Introduction}
Melda \cite{b20} is a delta-state conflict-free replicated data type (CRDT) \cite{b1,b10} designed for arbitrary JSON documents. It enables offline-first asynchronous collaboration, allowing multiple users to work on shared documents without needing a central coordination service. Melda supports various decentralized storage solutions, ensuring security, privacy, and data portability in collaborative document editing application. In contrast to operation-based CRDTs, such as AutoMerge \cite{b4} or Yjs \cite{b13}, Melda is based on delta states \cite{b3,b14}. Delta-state approaches address the major drawback of state-based replication, which is the potential for the state size to become very large. Delta state CRDTs operate by disseminating small updates, known as delta mutations. These updates are idempotent, meaning they can be applied multiple times to an existing state without compromising consistency, and can be transmitted over unreliable communication channels.
An important requirement that enables practical operation of such data structures is support for move operations, which allow for changing the location of data elements within a JSON structure. Move operations or relocations within a hierarchical structure like JSON are common in various use cases, such as reorganizing a document's structure, updating a filesystem-like structure, or modifying a nested configuration. However, supporting these operations can be challenging. On one hand, there's a need to produce an efficient representation of an update without serializing the entire state or copying the full contents of an object (and its children) to a new location and then deleting the old one. This approach would be inefficient and resource-intensive.
On the other hand, it's crucial to avoid unwanted results, such as duplicated nodes, where an object or subtree appears in both the old and new locations. This can lead to inconsistencies and errors in the data structure. To address these challenges, an effective solution must balance efficiency and consistency, ensuring that move operations are handled smoothly without compromising the integrity of the hierarchical structure.
The original version of Melda supported concurrent relocation (or reordering) of the objects inside an array by merging conflicting changes using a specific algorithm that would ensure that the elements would not be duplicated inside the resulting list. However concurrently moving objects inside the JSON hierarchy was not supported and would likely result in undesired outcomes, such as duplicate entries. In the following, we describe how we extended Melda to support move operations. The proposed enhancement involves minimal changes to the underlying data structure logic and does not introduce neither computational nor storage overhead when compared to the original solution.

\section{Melda}
In this section we will provide only a brief overview of Melda: for a more detailed analysis of the data structure and the design choices behind it refer to the original paper \cite{b20}. Melda aims to achieve eventual consistency by allowing each user to work on their own replica while offline and exchange the necessary information to update other replicas whenever possible. 
Melda is a CRDT based on a grow-only collection $C$ of JSON objects, replicated across multiple sites and concurrently updated by each participant. Objects are atomic and immutable, meaning even partial modifications replace the full content with a new version of the object, which is appended to the existing collection. Deletions are recorded using tombstones, a special value representing the final version of an object.
An object $o \in C$ is a data structure comprising zero or more name-value pairs. Each object is uniquely identified by a string value $id_o$, which remains unchanged even when the object is updated to a new version $o'$. To efficiently compare different versions of an object, the content $x$ is hashed to produce a string digest $H(x)$. The identifier of the object, $id_x$, is omitted from the hash computation, allowing different objects with the same content to have the same digest. We assume that the value associated to a specific version $x$ of an object can be efficiently retrieved using its digest $H(x)$.
To keep track of changes, each version is assigned a \textit{revision string}. All modifications made to an object are recorded as sequences of revision strings, however due to concurrent modifications, multiple sequences might exist for the same object. The history of modifications made to each object across all replicas is therefore recorded into a revision tree. For each revision tree an algorithm is used to deterministically compute the \textit{winning revision}, which typically corresponds to the latest version of the object. The collection of JSON objects $C$ and the set of all revision trees constitute the \textit{state} of the CRDT.

\begin{algorithm}
  \caption{Flattening Procedure}\label{algo:flattening_s1}
  \begin{algorithmic}[0]
    \State $C$ := \{\} \Comment{Associative array of extracted objects}
    \Procedure{Flatten}{$value$, $path$ := []}
	\Switch{type of $value$}
    	\Case{$String$}
      		\State \textbf{return} \Call{Escape}{$value$}
    	\EndCase
    	\Case{$Array$}
			\State $a\flat$ := [] \Comment{Flattened array}
    		\ForEach {$v \in value$}
    			\State $a\flat \gets a\flat$ $\cup$ \Call{Flatten}{$v$, $path$}
    		\EndFor
    		\State \textbf{return} $a\flat$
    	\EndCase
    	\Case{$Object$}
    		\State $o\flat$ := \{\} \Comment{Flattened object}
	    	\State $id_{o\flat}$ := \Call{MakeIdentifier}{$value$, $path$}
    		\ForEach {$[k,v] \in value$}
    			\State $o\flat[k] \gets $\Call{Flatten}{$v$,$path \cup [id_{o\flat},k]$}
    		\EndFor
    		\State $C[id_{o\flat}] \gets o\flat$
    		\State \textbf{return} $id_{o\flat}$
    	\EndCase
    	\Default{}
    		\textbf{return} $value$
    	\EndCase
  	\EndSwitch    
    \EndProcedure
  \end{algorithmic}
\end{algorithm}

To perform an update, Melda processes a JSON serialization of the application's data model, breaks it down into a collection of objects, and identifies changes by comparing these objects against the current state. Using a reversible transformation algorithm, the JSON document representing the data model is recursively flattened to move nested objects into an associative array $C$ (Algorithm \ref{algo:flattening_s1}). Although not shown in the pseudo-code, to allow for more control over the flattening procedure, the latter is enabled only when the key string ends with character $\flat$. Each moved object is replaced by a unique string reference generated by the \textsc{MakeIdentifier} function, which depends on either the value of an \textit{\_id} field or the object's path within the document. All other strings are escaped using an \textsc{Escape} function. The resulting object is designated as the \textit{root}. Subsequently, all objects in $C$ are compared against the current state and changes generate new revision. To minimize the space occupied by large arrays, a difference algorithm is used to create patches against previous revisions. To reconstruct the document from a given state, the object references in the \textit{root} object (which is the starting $value$) are recursively replaced by the value of the winning revision, as detailed in Algorithm \ref{algo:unflattening_s1}.

\begin{algorithm}[]
  \caption{Unflattening Procedure (original)}\label{algo:unflattening_s1}
  \begin{algorithmic}[0]
      \State $C$ := \{ ... \} \Comment{Associative array of extracted objects}
      \Procedure{Unflatten}{$value$}	
			\Switch{type of $value$}
    			\Case{$String$}
    				\If{\emph{value is an object identifier}}
                        \State $v := C[value]$
                        \State \textbf{return} \Call{Unflatten}($v$)
    				\Else
      				  \State \textbf{return} \Call{Unescape}{$value$}
                    \EndIf
      			\EndCase
      			\Case{$Array$}
      				\State $na$ := []
    				\ForEach {$v \in value$}
    					\State $na.append(\Call{Unflatten}{v})$
    				\EndFor
    				\State \textbf{return} $na$
    			\EndCase
      			\Case{$Object$}
    				\ForEach {$[key, v] \in value$}
    					\State $o[key] \gets \Call{Unflatten}{v}$
    				\EndFor
    				\State \textbf{return} $o$
    			\EndCase
      			\Default{}
    				\State \textbf{return} $value$
    			\EndCase
    		\EndSwitch
	  \EndProcedure
  \end{algorithmic}
\end{algorithm}

\subsection{Non-destructive array merging}
To correctly handle updates to arrays of objects, Melda implements a \textit{non-destructive array merging} algorithm which takes into account concurrent edits such as insertions of new objects. For example, consider concurrent updates on an array $[A,B,C]$ (where $A$, $B$, and $C$ are JSON objects). On the first replica, a new element \textbf{D} is appended at the end of the array (producing $[A,B,C,\textbf{D}]$), whereas on another replica a new element \textbf{E} is inserted between \textbf{A} and \textbf{B} (producing in $[A,\textbf{E},B,C]$). If we merge those CRDTs, we will end up with two conflicting revisions for the array, and either one of the two will be chosen. Therefore, the user might see either $[A,B,C,\textbf{D}]$ or $[A,\textbf{E},B,C]$. To retain both additions, Melda executes the algorithm listed in Algorithm \ref{algo:array_merging}, in order to produce a \textit{merged view} of the array. In the previous example, the merging procedure will output an array containing both the added object $[A,\textbf{E},B,C,\textbf{D}]$.
 
\begin{algorithm}[h]
  \caption{Array Merging Procedure}\label{algo:array_merging}
  \begin{algorithmic}[1]
      \Procedure{MergeArrays}{$S, T$}	
      \State $\iota := 0$
      \State $\pi := 0$
		 \ForEach {$e \in S$}
			\If{$e \in T$}
				\State $\iota \gets \Call{IndexOf}{T, e}$
				\State $break$
			\Else
			 	\State $\pi \gets \pi + 1$
			\EndIf
    	 \EndFor
    	 \State $\tau := 0$
    	 \ForEach {$e \in S$}
    	 	\If{$e \in T$}
    	 	 	\State $\iota \gets \Call{IndexOf}{T, e}$
    	 	\Else
    	 		\If{$\tau < \pi$}
    	 			\State $\Call{Insert}{T,\iota,e}$
    	 			\State $\pi \gets \tau$
    	 		\Else
    	 			\State $\iota \gets \iota + 1$
    	 			\State $\Call{Insert}{T,\iota,e}$
    	 		\EndIf
    	 	\EndIf
    	 	\State $\tau \gets \tau + 1$
    	 \EndFor
	  \EndProcedure
  \end{algorithmic}
\end{algorithm}

\section{Relocating objects}
In this paper we are concerned with relocating objects inside a JSON document. More specifically we are interested in efficient and consistent ways to deal with objects moving within the same container or to another container. In this regard, we consider the relation between the \textit{container} (which can be an object or an array) and the \textit{content} (which is an object) to be a parent-child one: the container being the \textit{parent}, whereas the object being the \textit{child}. Although there are different solutions to this problems, we are interested in merging concurrent edits involving relocations such that:
\begin{itemize}
    \item Moved objects are not duplicated;
    \item Moved objects do not disappear, unless one of the edits is a deletion;
    \item No cycles are created when merging concurrent moves.
    \item Moved objects are relocated to a new container consistent with at least one of the edits;
\end{itemize}

Before addressing move operations that relocate an object to a different level in the hierarchy, let's consider a simpler example where an object is moved without changing its parent, namely when the container does not change.

\subsection{Concurrent relocation without reparenting}
\begin{figure}[h]
  \begin{minipage}[c]{.25\textwidth}\centering
        \begin{tikzpicture} [level distance=10mm,
            every node/.style={circle,draw,inner sep=0pt,minimum size=7mm},
            level 1/.style={sibling distance=10mm},
            level 2/.style={sibling distance=10mm},
            level 3/.style={sibling distance=10mm}]
             \node (root) {$\sqrt{}$}
            child {
                node (nA) {A}
            }
            child {
                node (nB) {B}
            }
            child {
                node (nC) {C}
            };
        \end{tikzpicture}
    \end{minipage}\begin{minipage}[c]{.2\textwidth}
      \begin{lstlisting}[language=JSON]
{
    "A♭":{"_id":"A"},
    "B♭":{"_id":"B"},
    "C♭":{"_id":"C"}
}
\end{lstlisting}
    \end{minipage}
            \caption{Hierarchical structure with unordered children and corresponding JSON (version 0).}
        \label{fig:example_1}
\end{figure}
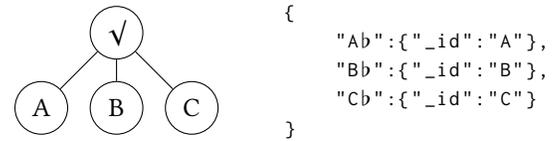

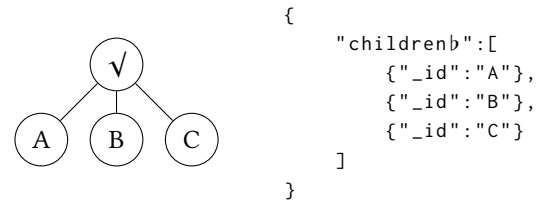
\begin{figure}[h]
  \begin{minipage}[c]{.25\textwidth}\centering
        \begin{tikzpicture} [level distance=10mm,
            every node/.style={circle,draw,inner sep=0pt,minimum size=7mm},
            level 1/.style={sibling distance=10mm},
            level 2/.style={sibling distance=10mm},
            level 3/.style={sibling distance=10mm}]
             \node (root) {$\sqrt{}$}
            child {
                node (nA) {A}
            }
            child {
                node (nB) {B}
            }
            child {
                node (nC) {C}
            };
        \end{tikzpicture}
    \end{minipage}\begin{minipage}[c]{.2\textwidth}\centering
      \begin{lstlisting}[language=JSON]
{
    "children♭":[
        {"_id":"A"},
        {"_id":"B"},
        {"_id":"C"}
    ]
}
\end{lstlisting}
    \end{minipage}
            \caption{Hierarchical structure with ordered children and corresponding JSON (version 0).}
        \label{fig:example_1a}
\end{figure}

Although this is a straightforward situation, there are two distinct ways, each with different characteristics, to represent a parent-child relationship in a JSON document.
Consider the example documents depicted in Figures \ref{fig:example_1} and \ref{fig:example_1a}. Each node represents a JSON object: the \textit{root} object, $A$, $B$, and $C$. Such a hierarchical structure can be either written with an unordered set (Figure \ref{fig:example_1}) or an ordered list (Figure \ref{fig:example_1a}) of children. In this example, $A$, $B$, and $C$ are children of the \textit{root} container object. If objects are referenced as \textit{fields} from within the \textit{root} object, children will be unordered; conversely, if objects are stored inside an array, their order is preserved.

\begin{figure}[h]
  \begin{minipage}[c]{.2\textwidth}\centering
        \begin{tikzpicture} [level distance=10mm,
            every node/.style={circle,draw,inner sep=0pt,minimum size=7mm},
            level 1/.style={sibling distance=10mm},
            level 2/.style={sibling distance=10mm},
            level 3/.style={sibling distance=10mm}]
             \node (root) {$\sqrt{}$}
            child {
                node (nB) {B}
            }
            child {
                node (nC) {C}
            };
        \end{tikzpicture}\\
        \small(version 1a)
    \end{minipage}\begin{minipage}[c]{.2\textwidth}\centering
        \begin{tikzpicture} [level distance=10mm,
            every node/.style={circle,draw,inner sep=0pt,minimum size=7mm},
            level 1/.style={sibling distance=10mm},
            level 2/.style={sibling distance=10mm},
            level 3/.style={sibling distance=10mm}]
             \node (root) {$\sqrt{}$}
            child {
                node (nD) {D}
            }
            child {
                node (nC) {C}
            }
            child {
                node (nB) {B}
            }
            child {
                node (nA) {A}
            };
        \end{tikzpicture}\\
        \small(version 1b)
    \end{minipage}
            \caption{Concurrent updates with creation and deletion but no reparenting.}
        \label{fig:example_1_complex}
\end{figure}
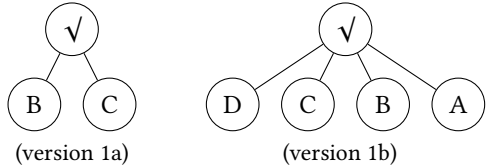

In the first case, \textit{moving} the relative position of the fields (keys) within an object (in this example the \textit{root} object) has no practical meaning, because fields are always ordered alphabetically. In the second case concurrent moves will be resolved consistently by the aforementioned \textit{non-destructive array merging} algorithm. 
A more complex scenario is illustrated in Figure \ref{fig:example_1_complex}: two concurrent updates either remove object $A$ (version 1a) or move object $A$ to the end and insert object $D$ at the beginning of the array (version 1b). 

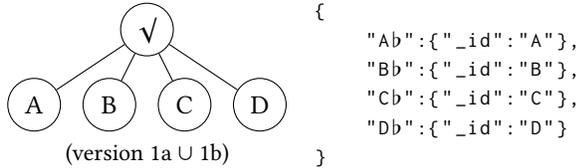
\begin{figure}[h]
  \begin{minipage}[c]{.25\textwidth}\centering
        \begin{tikzpicture} [level distance=10mm,
            every node/.style={circle,draw,inner sep=0pt,minimum size=7mm},
            level 1/.style={sibling distance=10mm},
            level 2/.style={sibling distance=10mm},
            level 3/.style={sibling distance=10mm}]
             \node (root) {$\sqrt{}$}
            child {
                node (nA) {A}
            }
            child {
                node (nB) {B}
            }
            child {
                node (nC) {C}
            }
            child {
                node (nD) {D}
            };
        \end{tikzpicture}\\
        \small(version 1a $\cup$ 1b)
    \end{minipage}\begin{minipage}[c]{.2\textwidth}\centering
      \begin{lstlisting}[language=JSON]
{
    "A♭":{"_id":"A"},
    "B♭":{"_id":"B"},
    "C♭":{"_id":"C"},
    "D♭":{"_id":"D"}
}
\end{lstlisting}
    \end{minipage}
            \caption{Possible structure after merge (unordered children): either one of the versions of the root object is retained.}
        \label{fig:example_merge_unordered}
\end{figure}

\begin{figure}[h]
  \begin{minipage}[c]{.25\textwidth}\centering
        \begin{tikzpicture} [level distance=10mm,
            every node/.style={circle,draw,inner sep=0pt,minimum size=7mm},
            level 1/.style={sibling distance=10mm},
            level 2/.style={sibling distance=10mm},
            level 3/.style={sibling distance=10mm}]
             \node (root) {$\sqrt{}$}
            child {
                node (nD) {D}
            }
            child {
                node (nC) {C}
            }
            child {
                node (nB) {B}
            };
        \end{tikzpicture}\\
        \small(version 1a $\cup$ 1b)
    \end{minipage}\begin{minipage}[c]{.2\textwidth}\centering
      \begin{lstlisting}[language=JSON]
{
    "children♭":[
        {"_id":"D"},
        {"_id":"C"},
        {"_id":"B"}
    ]
}
\end{lstlisting}
    \end{minipage}
            \caption{Structure after merge (ordered children): changes made to the root object's array are merged.}
        \label{fig:example_merge_ordered}
\end{figure}
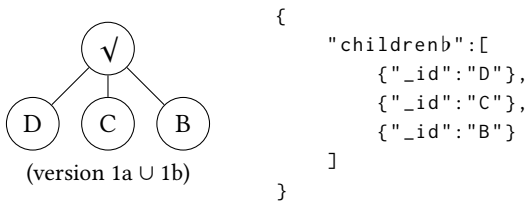

As shown in Figure \ref{fig:example_merge_unordered}, with an unordered hierarchy either version 1a or 1b is chosen after merging. However, with the ordered hierarchy (Figure \ref{fig:example_merge_ordered}) the array merging algorithm ensures that changes made in both versions are retained, and the result is consistent with the edits: object $A$ has been deleted (as in version 1a), object $C$ is moved before $B$, and $D$ has been added to the front of the array (as in version 1b).
With the original version of Melda\cite{b20} it is therefore possible to safely relocate objects inside the JSON document, provided that all objects \textit{remain} within the same array. More specifically, using a \textit{non-destructive array merging} algorithm, it is possible to merge different versions of an array (produced by concurrent moves) into a single coherent view without duplicates.

\subsection{Concurrent relocation with reparenting}
\begin{figure}[h]
\begin{minipage}[c]{.15\textwidth}\centering
        \begin{tikzpicture} [level distance=10mm,
            every node/.style={circle,draw,inner sep=0pt,minimum size=7mm},
            level 1/.style={sibling distance=10mm},
            level 2/.style={sibling distance=10mm},
            level 3/.style={sibling distance=10mm}]
             \node (root) {$\sqrt{}$}
             child {
                node (nA) {A}
                child {
                    node (nC) {C}
                }
            }
            child {
                node (nB) {B}
            };
        \end{tikzpicture}\\
        \small(version 0$\star$)
    \end{minipage}
  \begin{minipage}[c]{.15\textwidth}\centering
        \begin{tikzpicture} [level distance=10mm,
            every node/.style={circle,draw,inner sep=0pt,minimum size=7mm},
            level 1/.style={sibling distance=10mm},
            level 2/.style={sibling distance=10mm},
            level 3/.style={sibling distance=10mm}]
             \node (root) {$\sqrt{}$}
            child {
                node (nA) {A}
                child {
                    node (nB) {B}
                }
                child {
                    node (nC) {C}
                }
            };
        \end{tikzpicture}\\
        \small(version 1a$\star$)
    \end{minipage}\begin{minipage}[c]{.15\textwidth}\centering
        \begin{tikzpicture} [level distance=10mm,
            every node/.style={circle,draw,inner sep=0pt,minimum size=7mm},
            level 1/.style={sibling distance=10mm},
            level 2/.style={sibling distance=10mm},
            level 3/.style={sibling distance=10mm}]
             \node (root) {$\sqrt{}$}
            child {
                node (nB) {B}
                child {
                    node (nA) {A}
                    child {
                        node (nC) {C}
                    }
                }
            };
        \end{tikzpicture}\\
        \small(version 1b$\star$)
    \end{minipage}
            \caption{Two different updates with concurrent moves to a different level of the hierarchy.}
        \label{fig:example_1_updates}
\end{figure}
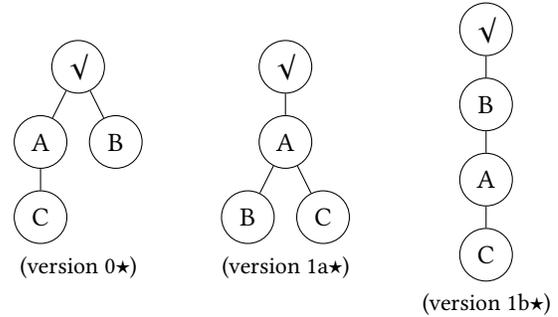

As we have hinted at the beginning of this paper, the original algorithm is not able to properly deal with problems arising from relocation of objects inside the JSON hierarchy that change the parent of an object. To understand the potential issues that can affect the outcome of concurrent edits we will first consider the example depicted in Figure \ref{fig:example_1_updates} (adapted from \cite{b21}), which shows two concurrent updates performed on the initial data structure. For simplicity we only consider the case of an ordered hierarchy (i.e. children objects are stored inside an array in the parent object).

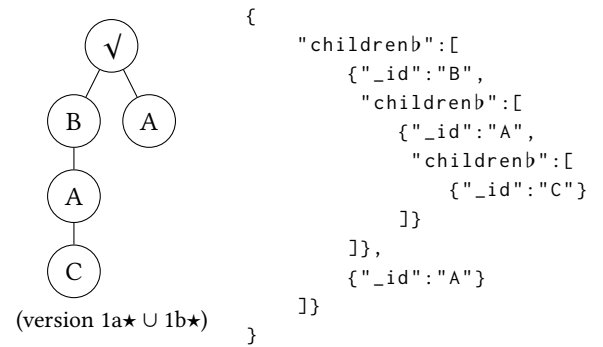
\begin{figure}[h]
  \begin{minipage}[c]{.2\textwidth}\centering
        \begin{tikzpicture} [level distance=10mm,
            every node/.style={circle,draw,inner sep=0pt,minimum size=7mm},
            level 1/.style={sibling distance=10mm},
            level 2/.style={sibling distance=10mm},
            level 3/.style={sibling distance=10mm}]
             \node (root) {$\sqrt{}$}
            child {
                node (nB) {B}
                child {
                    node (nA) {A}
                    child {
                        node (nC) {C}
                    }
                }
            }
            child {
                node (nA) {A}
            };
        \end{tikzpicture}\\
        \small(version 1a$\star$ $\cup$ 1b$\star$)
    \end{minipage}\begin{minipage}[c]{.2\textwidth}\centering
      \begin{lstlisting}[language=JSON]
{
    "children♭":[
        {"_id":"B",
         "children♭":[
            {"_id":"A",
             "children♭":[
                {"_id":"C"}
            ]}
        ]},
        {"_id":"A"}
    ]}
}
\end{lstlisting}
    \end{minipage}
            \caption{Structure after merge (ordered children) using the original Melda: notice how object $A$ is duplicated.}
        \label{fig:wrong_merge}
\end{figure}

With the original version of Melda, those edits would result in a merged document where object $A$ is duplicated (Figure \ref{fig:wrong_merge}), violating our initial requirements.

\subsection{Concurrent relocation (with reparenting) and deletion}
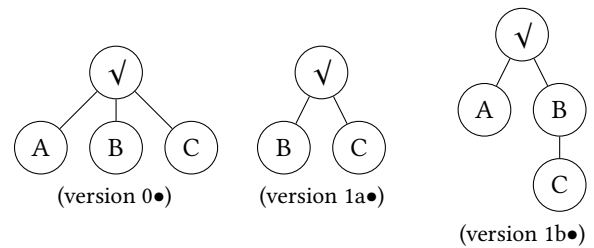
\begin{figure}[h]
\begin{minipage}[c]{.15\textwidth}\centering
        \begin{tikzpicture} [level distance=10mm,
            every node/.style={circle,draw,inner sep=0pt,minimum size=7mm},
            level 1/.style={sibling distance=10mm},
            level 2/.style={sibling distance=10mm},
            level 3/.style={sibling distance=10mm}]
             \node (root) {$\sqrt{}$}
            child {
                node (nA) {A}
            }
            child {
                node (nB) {B}
            }
            child {
                node (nC) {C}
            };
        \end{tikzpicture}\\
        \small(version 0$\bullet$)
    \end{minipage}
  \begin{minipage}[c]{.15\textwidth}\centering
        \begin{tikzpicture} [level distance=10mm,
            every node/.style={circle,draw,inner sep=0pt,minimum size=7mm},
            level 1/.style={sibling distance=10mm},
            level 2/.style={sibling distance=10mm},
            level 3/.style={sibling distance=10mm}]
             \node (root) {$\sqrt{}$}
                child {
                    node (nB) {B}
                }
                child {
                    node (nC) {C}
                };
        \end{tikzpicture}\\
        \small(version 1a$\bullet$)
    \end{minipage}\begin{minipage}[c]{.15\textwidth}\centering
        \begin{tikzpicture} [level distance=10mm,
            every node/.style={circle,draw,inner sep=0pt,minimum size=7mm},
            level 1/.style={sibling distance=10mm},
            level 2/.style={sibling distance=10mm},
            level 3/.style={sibling distance=10mm}]
             \node (root) {$\sqrt{}$}
            child {
                node (nA) {A}
            }
            child {
                node (nB) {B}
                child {
                        node (nC) {C}
                }
            };
        \end{tikzpicture}\\
        \small(version 1b$\bullet$)
    \end{minipage}
            \caption{Two different updates with a concurrent move and a deletion.}
        \label{fig:example_2_updates}
\end{figure}

We now consider an even more complex example, involving both concurrent relocation (with reparenting) and deletion (Figure \ref{fig:example_2_updates}). The initial structure (version 0$\bullet$) is concurrently update by removing object $A$ (version 1a$\bullet$), or by moving object $C$ inside object $B$ (version 1b$\bullet$). The tombstone object is not depicted because it is automatically ignored when reconstructing the array.

\begin{figure}[h]
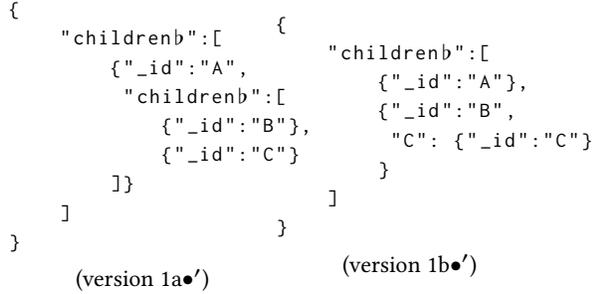

  \begin{minipage}[c]{.2\textwidth}\centering
      \begin{lstlisting}[language=JSON]
{
    "children♭":[
        {"_id":"A",
         "children♭":[
            {"_id":"B"},
            {"_id":"C"}
        ]}
    ]
}
\end{lstlisting}
\small(version 1a$\bullet'$)
    \end{minipage}\begin{minipage}[c]{.2\textwidth}\centering
      \begin{lstlisting}[language=JSON]
{
    "children♭":[
        {"_id":"A"},
        {"_id":"B",
         "C": {"_id":"C"}
        }
    ]
}
\end{lstlisting}
\small(version 1b$\bullet'$)
    \end{minipage}
            \caption{Relocation between two different data structures: object $C$ moves from an array to a field.}
        \label{fig:relocation_mix}
\end{figure}

To make things more challenging, for this scenario we also consider both a relocation inside another array and a relocation that moves the object inside a field of another object. Figure \ref{fig:relocation_mix} lists the these two new types of concurrent updates to version 0$\bullet'$: notice how object $C$ remains within the children's array of $A$ in version 1a$\bullet'$, while being relocated as a field of object $B$ in version 1b$\bullet'$.

\begin{figure}[h]
  \begin{minipage}[c]{.2\textwidth}\centering
        \begin{tikzpicture} [level distance=10mm,
            every node/.style={circle,draw,inner sep=0pt,minimum size=7mm},
            level 1/.style={sibling distance=10mm},
            level 2/.style={sibling distance=10mm}, 
            level 3/.style={sibling distance=10mm}]
             \node (root) {$\sqrt{}$}
            child {
                node (nB) {B}
                child {
                        node (nC) {C}
                }
                child {
                        node (nC) {C}
                }            }
            ;
        \end{tikzpicture}\\
        \small(version 1a$\bullet$ $\cup$ 1b$\bullet$)
    \end{minipage}\begin{minipage}[c]{.2\textwidth}\centering
      \begin{lstlisting}[language=JSON]
{
    "children♭":[
        {"_id":"B",
         "children♭":[
            {"_id":"C"},
            {"_id":"C"}
        ]}
    ]}
}
\end{lstlisting}
    \end{minipage}
            \caption{Structure after merge (ordered children) using the original Melda: notice how object $C$ is duplicated.}
        \label{fig:wrong_merge_2}
\end{figure}
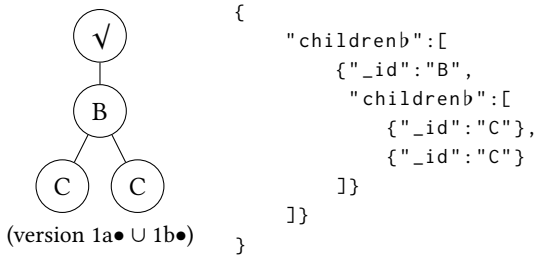

Again, the original Melda algorithm fails to correctly merge these concurrent edits, and produces a resulting JSON document with a duplicated object $C$ (Figure \ref{fig:wrong_merge_2}). The merge of the data structures presented in Figure \ref{fig:relocation_mix}, although not shown, also results in a duplicated $C$ object.

\section{Introducing support for move operations}
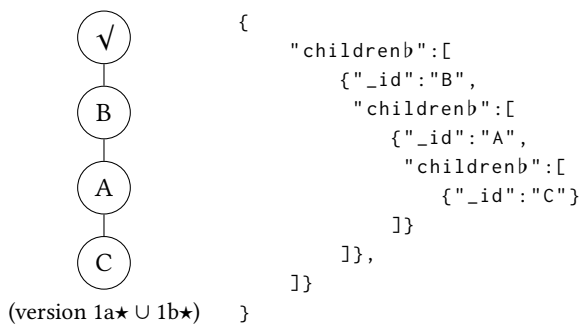
\begin{figure}[h]
  \begin{minipage}[c]{.2\textwidth}\centering
        \begin{tikzpicture} [level distance=10mm,
            every node/.style={circle,draw,inner sep=0pt,minimum size=7mm},
            level 1/.style={sibling distance=10mm},
            level 2/.style={sibling distance=10mm},
            level 3/.style={sibling distance=10mm}]
             \node (root) {$\sqrt{}$}
            child {
                node (nB) {B}
                child {
                    node (nA) {A}
                    child {
                        node (nC) {C}
                    }
                }
            };
        \end{tikzpicture}\\
        \small(version 1a$\star$ $\cup$ 1b$\star$)
    \end{minipage}\begin{minipage}[c]{.2\textwidth}\centering
      \begin{lstlisting}[language=JSON]
{
    "children♭":[
        {"_id":"B",
         "children♭":[
            {"_id":"A",
             "children♭":[
                {"_id":"C"}
            ]}
        ]},
    ]}
}
\end{lstlisting}
    \end{minipage}
            \caption{Structure after merge (ordered children) using the updated Melda: notice how the improved algorithm produces a valid result.}
        \label{fig:good_merge}
\end{figure}

\begin{figure}[h]
  \begin{minipage}[c]{.2\textwidth}\centering
        \begin{tikzpicture} [level distance=10mm,
            every node/.style={circle,draw,inner sep=0pt,minimum size=7mm},
            level 1/.style={sibling distance=10mm},
            level 2/.style={sibling distance=10mm},
            level 3/.style={sibling distance=10mm}]
             \node (root) {$\sqrt{}$}
            child {
                node (nB) {B}
                child {
                        node (nC) {C}
                }
                }
            ;
        \end{tikzpicture}\\
        \small(version 1a$\bullet'$ $\cup$ 1b$\bullet'$)
    \end{minipage}\begin{minipage}[c]{.2\textwidth}\centering
      \begin{lstlisting}[language=JSON]
{
    "children♭":[
        {"_id":"B",
         "C♭":{"_id":"C"}
        }
    ]
}
\end{lstlisting}
    \end{minipage}
            \caption{Structure after merge using the updated Melda: the result now meets the requirements even when an object is moved between an array and another object.}
        \label{fig:good_merge_3}
\end{figure}
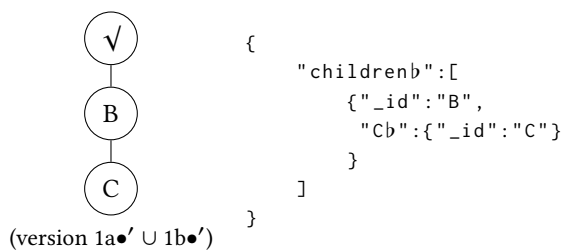

As shown in the previous examples, the original version of Melda fails at merging JSON documents with concurrent moves that involve reparenting. We investigated this issue and the fix (or improvement) turned out to be quite simple. To avoid duplicates the unflattening procedure was changed so that an object can only be \textit{used} once while reconstructing the document. Accordingly, we simply remove each referenced object from the associative array $C$, and ignore situations where a reference points to an object that is missing in the same data structure (Algorithm \ref{algo:unflattening_s2}). We understand that this solution \textit{favors} containers which are reconstructed first in the document, however it also ensures that the merged document cannot contain duplicate objects. Furthermore, since the unflattening procedure reconstructs the hierarchy recursively (using a \textit{depth-first}) approach, we ensure that no cycle can be ever created. By running the improved algorithm on the previously considered examples we now obtain results that satisfy our requirements (Figures \ref{fig:good_merge}, \ref{fig:good_merge_2}, and \ref{fig:good_merge_3}). Concurrent edits which involve moving objects to a different location in the document do not result in duplications. Concerning the algorithmic complexity, the introduced change does not significantly affect the estimations provided in \cite{b20}. Moreover, only the \textit{read} operation is concerned (because the unflattening procedure is only applied there).
Since the improvement is only about the logic and does not require storing additional information, the performance and storage overhead are practically unchanged when compared to the original version. 

\begin{algorithm}[]
  \caption{Unflattening Procedure (updated)}\label{algo:unflattening_s2}
  \begin{algorithmic}[0]
      \State $C$ := \{ ... \} \Comment{Associative array of extracted objects}
      \Procedure{Unflatten}{$value$}	
			\Switch{type of $value$}
    			\Case{$String$}
    				\If{\emph{value is an object identifier}}
                        \If{$value \in C$}
                            \State $v := C[value]$
                            \State \textbf{$C \gets C \setminus value$}
                            \State \textbf{return} $Unflatten(v)$
                        \EndIf
    				\Else
      				  \State \textbf{return} \Call{Unescape}{$value$}
                    \EndIf
      			\EndCase
      			\Case{$Array$}
      				\State $na$ := []
    				\ForEach {$v \in value$}
    					\State $na.append(\Call{Unflatten}{v})$
    				\EndFor
    				\State \textbf{return} $na$
    			\EndCase
      			\Case{$Object$}
    				\ForEach {$[key, v] \in value$}
    					\State $o[key] \gets \Call{Unflatten}{v}$
    				\EndFor
    				\State \textbf{return} $o$
    			\EndCase
      			\Default{}
    				\State \textbf{return} $value$
    			\EndCase
    		\EndSwitch
	  \EndProcedure
  \end{algorithmic}
\end{algorithm}

\begin{figure}[h]
  \begin{minipage}[c]{.2\textwidth}\centering
        \begin{tikzpicture} [level distance=10mm,
            every node/.style={circle,draw,inner sep=0pt,minimum size=7mm},
            level 1/.style={sibling distance=10mm},
            level 2/.style={sibling distance=10mm},
            level 3/.style={sibling distance=10mm}]
             \node (root) {$\sqrt{}$}
            child {
                node (nB) {B}
                child {
                        node (nC) {C}
                }
                }
            ;
        \end{tikzpicture}\\
        \small(version 1a$\bullet$ $\cup$ 1b$\bullet$)
    \end{minipage}\begin{minipage}[c]{.2\textwidth}\centering
      \begin{lstlisting}[language=JSON]
{
    "children♭":[
        {"_id":"B",
         "children♭":[
            {"_id":"C"}
        ]}
    ]}
}
\end{lstlisting}
    \end{minipage}
            \caption{Structure after merge (ordered children) using the updated Melda: the result now meets the requirements (no duplications).}
        \label{fig:good_merge_2}
\end{figure}
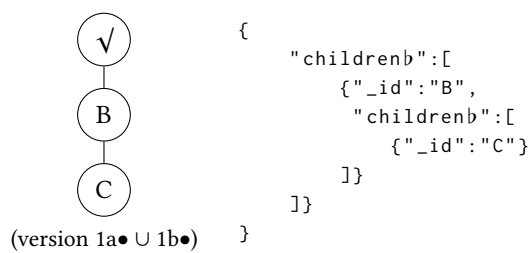

\section{Evaluation}
\begin{figure}[h]
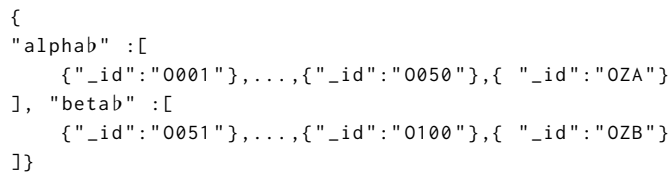

\begin{lstlisting}[language=JSON]
{
"alpha♭" :[
    {"_id":"O001"},...,{"_id":"O050"},{ "_id":"OZA"}
], "beta♭" :[
    {"_id":"O051"},...,{"_id":"O100"},{ "_id":"OZB"}
]}
\end{lstlisting}
\caption{Example document used during evaluation.}
        \label{fig:evaluation}
\end{figure}

The proposed modification was implemented in Melda \footnote{{\url{https://github.com/slashdotted/libmelda}}} to 
evaluate the correctness of the solution with respect to the requirements, namely that moved objects are not duplicated, that there are no cycles in the resulting structure, and that objects still present after each merge (unless deleted by one of the edits). We considered a scenario with $10$ clients performing concurrent updates to the starting document listed in Figure \ref{fig:evaluation}. The document contains two arrays with $51$ objects each. Each edit moves several random objects between the $alpha$ and $beta$ arrays. A total of $100$ edits where concurrently performed by each client. After each edit it was verified that the resulting document contained no repeated objects and no object was missing. Furthermore, at each step it was verified that the merged document also satisfied those requirements.

\section{Related Work}
Support for move operations in a JSON CRDT has been fist proposed in \cite{b21}, for use within the Automerge library \cite{b4}. Compared to the solution presented in this paper, and due to the inherent differences in the underlying data model, Automerge requires a specific move operation in order to record the element being moved and the identifier of the destination (either a field inside a target object or an element inside a list). This makes the Automerge approach more complex compared to Melda. Other CRDTs for JSON documents, such as \cite{b13}, do not unfortunately provide support for move operations.

\section{Conclusions}
In this paper, we presented an extension to Melda\cite{b20} to support move operations. This enhancement relies on minimal changes to the underlying logic of the data structure, ensuring simplicity while addressing multiple use cases. This improvement addresses previous limitations, such as the presence of duplicate objects when relocating hierarchical elements. The solution ensures consistency by effectively managing conflicting edits involving moved objects. Overall, this enhancement makes Melda a more robust and versatile tool for collaborative applications that require efficient and consistent handling of hierarchical data. Future work will focus on further benchmarking and comparison with other approaches, and formally evaluating the correctness of our solution (in particular by determining corner cases).

\bibliographystyle{acm}
\bibliography{meldamove-2025-brocco}

\end{document}